\documentclass[mathleft
]{an}
\usepackage{graphicx}
\usepackage{times}
\def\approxgt{\mathrel{\hbox{\rlap{\lower.55ex \hbox {$\sim$}}
        \kern-.3em \raise.4ex \hbox{$>$}}}}
\def\approxlt{\mathrel{\hbox{\rlap{\lower.55ex \hbox {$\sim$}}
        \kern-.3em \raise.4ex \hbox{$<$}}}}

\begin{document}

\Pagespan{789}{}
\Yearpublication{2006}%
\Yearsubmission{2005}%
\Month{11}%
\Volume{999}%
\Issue{88}%

\title{Statistics of relativistically broadened Fe K$_{\alpha}$ lines in AGN}

\author{Matteo Guainazzi\inst{1}, \fnmsep\thanks{Corresponding author:
  Matteo Guainazzi.
  \email{Matteo.Guainazzi@sciops.esa.int}\newline}
Stefano Bianchi\inst{1}, Michal Dov\v ciak\inst{2}
}
\titlerunning{Statistics of relativistically broadened iron lines in AGN}
\authorrunning{M.Guainazzi, S.Bianchi \& M.Dov\v ciak}
\institute{
European Space Astronomy Centre of ESA, Apartado 50727, E-28080 Madrid, Spain
\and 
Astronomical Institute, Academy of Sciences of the Czech Republic, Bo\v cn\'i II 1401, 14131 Prague, Czech Republic
}

\received{30 August 2006}
\accepted{}
\publonline{}

\keywords{quasars: emission lines --
          galaxies: nuclei --
          galaxies: active --
          X-ray: galaxies --
          line: profiles}

\abstract{We present preliminary results on the properties of
relativistically broadened Fe K$_{\alpha}$ lines in a
collection of more then 100 Active Galactic Nuclei (AGN) observed
by the XMM-Newton EPIC-pn camera. Our main conclusions can
be summarized as follows: a) we detect broad lines in about
25\% of the sample objects. This fraction increases
to $42 \pm 13\%$ if we consider only objects with more
than $10^4$ counts in the hard (2--10~keV) band, and
to $50 \pm 32\%$ for the small sub-sample (6 objects) of
type~1 Piccinotti AGN 
with optimal XMM-Newton exposure (at least
$2 \times 10^5$ counts in the hard band); b) we find no significant difference
in the detection rate of broad lines between
obscured and unobscured AGN; c) the strongest relativistic profiles are measured
in low-luminosity ($L_X < 10^{43}$~erg~s$^{-1}$) AGN;
d) Equivalent Widths (EWs)
associated with relativistic profiles in stacked spectra
are $\approxlt$150~eV for all luminosity classes;
e) models of relativistically broadened
iron line profiles ({\tt kyrline}, \cite{dovciak05}),
which include full relativistic treatment 
of the accretion disk emission around a
Kerr black hole in the strong gravity regime, yield
an average disk inclination angle $\simeq$30$^{\circ}$,
and a 
radial dependence of the disk emissivity profile $\simeq$-3.
The distribution of EW is very broad, with
$\langle \log (EW) \rangle = 2.4$ and
$\sigma_{\log (EW)} = 1.4$. We estimate that
an investment of about 1~Ms of XMM-Newton
time would be required to put these results on
a sound statistical basis.}

\maketitle

\section{Introduction}

\begin{figure}
\includegraphics[width=80mm,height=70mm]{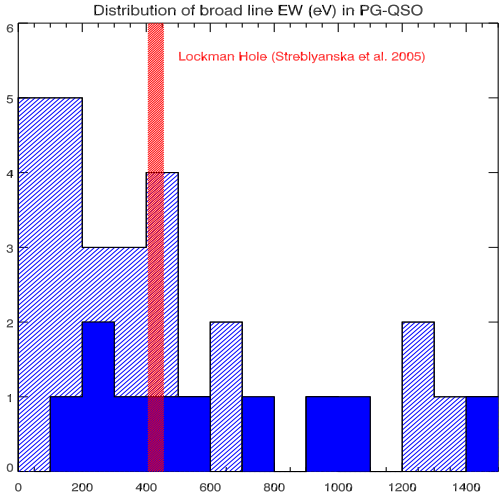}
\caption{Distribution of the Fe K$_{\alpha}$
line broad component EW
in the sample of 40 PG quasars
studied by Jim\'enez-Bail\'on et al. (2005).
{\it Shaded cells} represent 90\% confidence level upper limits.
The measured EW in Lockman Hole unobscured AGN [after
Tab.~1 in Streblyanska et
al. (2005)] is marked by the {\it hatched rectangle}
[$420 \pm^{35}_{30}$~eV when the broad line
is modeled with a Gaussian profile, as
in Jim\'enez-Bail\'on et al. (2005)].}
\label{fig1}
\end{figure}

XMM-Newton is the most suitable mission to study the properties
of broad iron K$_{\alpha}$ emission lines, thanks to the unprecedented collecting
area of its EPIC cameras. As this workshop has demonstrated, the
observational properties of the Fe emission line
complex have been one of the primary subjects of AGN
studies with XMM-Newton (see Fabian, this volume, and references therein).
Although most of the papers in this field have focused on the
detailed properties of the iron line profile in individual bright AGN,
there have appeared in the literature some attempts to estimate
the statistical importance of relativistic effects.
Stacking spectra of a large sample of $z \sim 1$
AGN in the Lockman Hole, for instance, Streblyanska et
al. (2005) unveiled a relativistically broadened and skewed profile in
both type~1 and type~2 objects. The Equivalent Width, EW,
of such a feature is large
(in the range 420-560~eV for unobscured, 280-460~eV for obscured AGN,
depending on the profile models used).
However, broadened iron K$_{\alpha}$ lines
with such a large EW are uncommon in the local universe
(cf., {\it e.g.}, the distribution of the broad Fe line EWs in a sample
of 40 PG quasars studied by Jim\'enez-Bail\'on et al. 2005 in Fig.~\ref{fig1}).

After almost seven years of successful XMM-Newton operations, it's time
to give a systematic look backward at the wealth of data contained in the
archive. This is the approach that has inspired the study presented in this
paper. We aim at answering primarily the following questions:

\begin{itemize}

\item {\it how often are effects due to X-ray illumination of a
relativistic accretion disk detected in AGN?}

\item {\it what do the properties of relativistically broadened profiles tell
us about the accretion flow in the innermost regions around
super-massive black holes?}

\end{itemize}

\subsection{The sample}

The sample comprises 102 AGN observed as pointed nominal targets
by XMM-Newton, and whose data were available in the public archive as of
June 2006. Out of them: 32 are classified as Seyfert~1s, 18 as Narrow
Line Seyfert~1s (NLSy1s), 39 are Radio Quiet Quasars (RQQs),
and 13 are Seyfert~2s.
No attempt has been made to create a complete and unbiased
sample.
We have restricted our study to Compton-thin obscured Seyfert~2s, whose
nuclear emission is covered by a column density $N_H \le 10^{22.5}$~cm$^{-2}$,
because the determination of the K$_{\alpha}$ iron line profile in
spectra affected by stronger obscuration is uncertain.
Most of the objects are local, with $\simeq$60\% of them having a
redshift $< 0.1$, and 98\% of them $<0.5$.

\section{Data reduction and analysis}

For each of the observations, data have been reduced
from the {\it Observation Data Files} using SASv6.5
(\cite{gabriel03}), and the most advanced calibration files available as of
March 2006. We present in this paper data from the
pn camera (\cite{struder01}) only, given its larger
effective area.
Spectra were extracted using standard
pattern event
selection. The radius of the circular spectral extraction regions and
the count rate background rejection thresholds were chosen in order
to maximize the signal-to-noise in each individual spectrum. Background spectra
were extracted from nearby circular areas in the same chip as the source,
except for observations in Small Window mode, where blank field event list
(\cite{read03})
were used, after coordinate recasting and normalization to the
quiescent background level. For the time being, we have not
merged different observations of the same target. Whenever multiple
observations are available, we have used that with the longest ``good''
exposure time (except in Fig.~\ref{fig9}, see Sect.~3).

All the spectra have been rebinned, in order to ensure that each
background-subtracted spectral channel has at least 25 counts, and that
the instrumental energy resolution is not oversampled by a factor larger
than 3. Spectra have been fit in the 2.5-15~keV energy range. We have
employed the following baseline fitting model:
\begin{equation}
\label{eqn1}
e^{- \sigma_\mathrm{ph} N_{\mathrm H}} \times A[E^{-\Gamma} + C(\Gamma,\phi, E) + \Sigma_{i=1}^3 G_i + B(\phi, \beta, a)]
\end{equation}
where $\sigma_{ph}$ is the photoelectric cross-section (\cite{anders89}),
$N_H$ is the column density (constrained not to be lower than the contribution
due to intervening gas in our Galaxy), $C$ is a Compton-reflection component
by a plane-parallel
neutral slab (model {\tt pexrav} in {\sc Xspec}; \cite{magdziarz95}),
$G_i$ are Gaussian unresolved profiles with centroid energy 6.4~keV
(Fe{\sc i}), 6.7~keV (Fe{\sc xxv}), and 6.96~keV (Fe{\sc xxvi}), and $B$
is a relativistically broadened profile. We have used for $B$
model profiles as predicted around Schwarzschild ($a = 0$;
\cite{fabian89}) and
Kerr ($a \simeq 1$; \cite{laor91}) black holes,
as well a set of models ({\tt kyrline}; \cite{dovciak05}),
which includes full relativistic treatment 
of the accretion disk emission around a
Kerr black hole in the strong gravity regime.
The {\tt ky} model family depends on the disk inclination
angle $\phi$, on the radial dependence of the emissivity (power-law index
$\beta$) and on the dimensionless
black hole spin $a$. In order not to over-fit the data in
moderate statistical quality spectra, we have employed the $\beta$
parameter only as a measurement of the strength of relativistic effects,
while
fixing the innermost and outermost radii of the line-emitting region to
the innermost stable circular orbit for a given $a$ value, and to
400 gravitational radii, respectively.

\section{Detection of relativistic lines in individual spectra}

\begin{figure}
\hbox{
\hspace{-0.5cm}
\includegraphics[width=85mm,height=75mm]{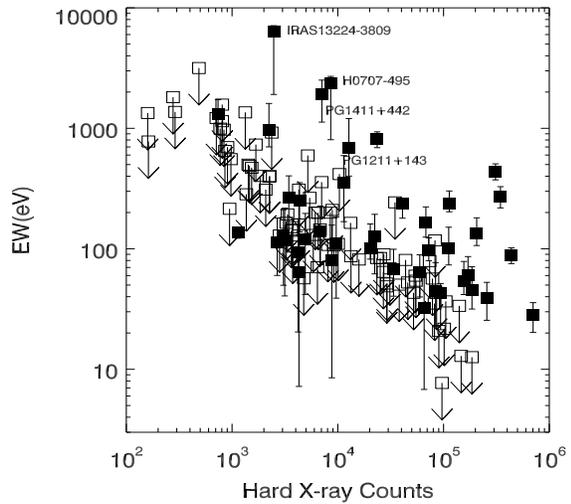}
}
\caption{EW of the relativistically broadened Fe K$_{\alpha}$ line
against the net hard ({\it i.e.} 2--10~keV) counts.
Each data point corresponds to one XMM-Newton
observation. Multiple observations of the same objects are
shown as distinct points.
{\it Filled
squares} represent line detections, {\it empty squares}
represent upper limits. Source excluded from the sample are
labeled (see text for details). The three objects
with EW upper limits $<$20~eV are: IC4329A (see, {\it e.g.}
\cite{gondoin01,steenbrugge05}), Mkn~509 (\cite{pounds01}), NGC7469
(\cite{blustin03}).}
\label{fig9}
\end{figure}

In Fig.~\ref{fig9} we show the EW of the Fe K$_{\alpha}$ relativistically
broadened component as a function of the total net counts in the 2--10~keV
band. As expected, the overwhelming majority of the measurements are aligned
along a straight line (in log-log space), which represents the sensitivity
limit of our sample. Three
poorly exposed ($<$10~000 hard counts)
objects are well above this correlation,
and exhibit EW by an order of magnitude larger than the average of the sample
at their net count level. They are: IRAS~13224-3809, PG1411+442, and 1H~0707-495. 
In the last the presence of a $EW > 1$~keV
relativistically broadened Fe K$_{\alpha}$ line
has been challenged 
(\cite{gallo04,fabian04}). We therefore consider
these detections as ``suspicious'', and removed these
objects from the
sample. For similar reason, we exclude from the
sample PG1211+143 (\cite{pounds03}).
None of the sample results presented in this paper is
significantly affected
by this choice.

\begin{figure}
\includegraphics[width=80mm,height=70mm]{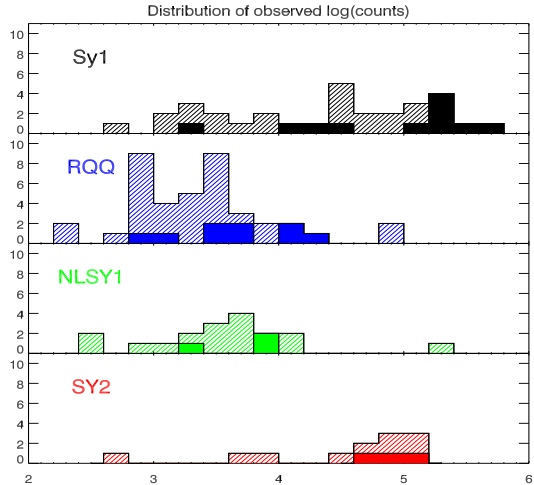}
\caption{Distribution of the net counts in the hard X-ray band
for the AGN of our sample. {\it Filled cells}
correspond to detections of a broad iron K$_{\alpha}$ profile;
{\it shaded cells} to non-detections.}
\label{fig2}
\end{figure}

\begin{figure}
\hbox{
\hspace{-1.5cm}
\includegraphics[width=95mm,height=75mm]{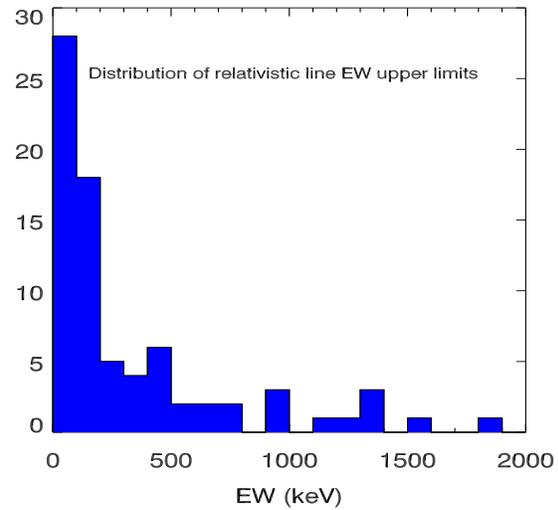}
}
\caption{Distribution of upper limits on the EW of the broad
iron K$_{\alpha}$ line in our sample.}
\label{fig3}
\end{figure}

The distribution of the source counts is shown in Fig.~\ref{fig2}. In
these histograms, {\it filled cells} represent objects where a broad Fe
K$_{\alpha}$ iron line is detected, {\it shaded cells} represent
non-detections (the distribution of the EW upper limits is shown in
Fig.~\ref{fig3}). The fraction of detections is $42 \pm 13\%$ and $16 \pm 6\%$
for ``well-exposed'', and ``under-exposed'' objects, respectively, where we
have arbitrarily set a threshold of 10000 net counts to distinguish the
two classes. The detection fraction is not significantly dependent on the
AGN type (Tab.~\ref{tab1}); however, the small number statistics
\begin{table}
\begin{center}
\begin{tabular}{lcc} \hline \hline
AGN type & Well exposed & Underexposed \\
& (\%) & (\%) \\ \hline
Seyfert~1s & $50 \pm 15$ & $<21$ \\
RQQs & $60 \pm 48$ & $17 \pm 8$ \\
NLSy1~s & $<30$ & $20 \pm 14$ \\
Seyfert~2s & $33 \pm 24$ & $<30$ \\ \hline \hline
\end{tabular}
\end{center}
\caption{Fraction of broad iron K$_{\alpha}$ detections
in our AGN sample. Errors are at the 1$\sigma$ level}
\label{tab1}
\end{table}
prevents strong statement from being drawn.

\begin{figure}
\hbox{
\hspace{-1.0cm}
\includegraphics[width=90mm,height=70mm]{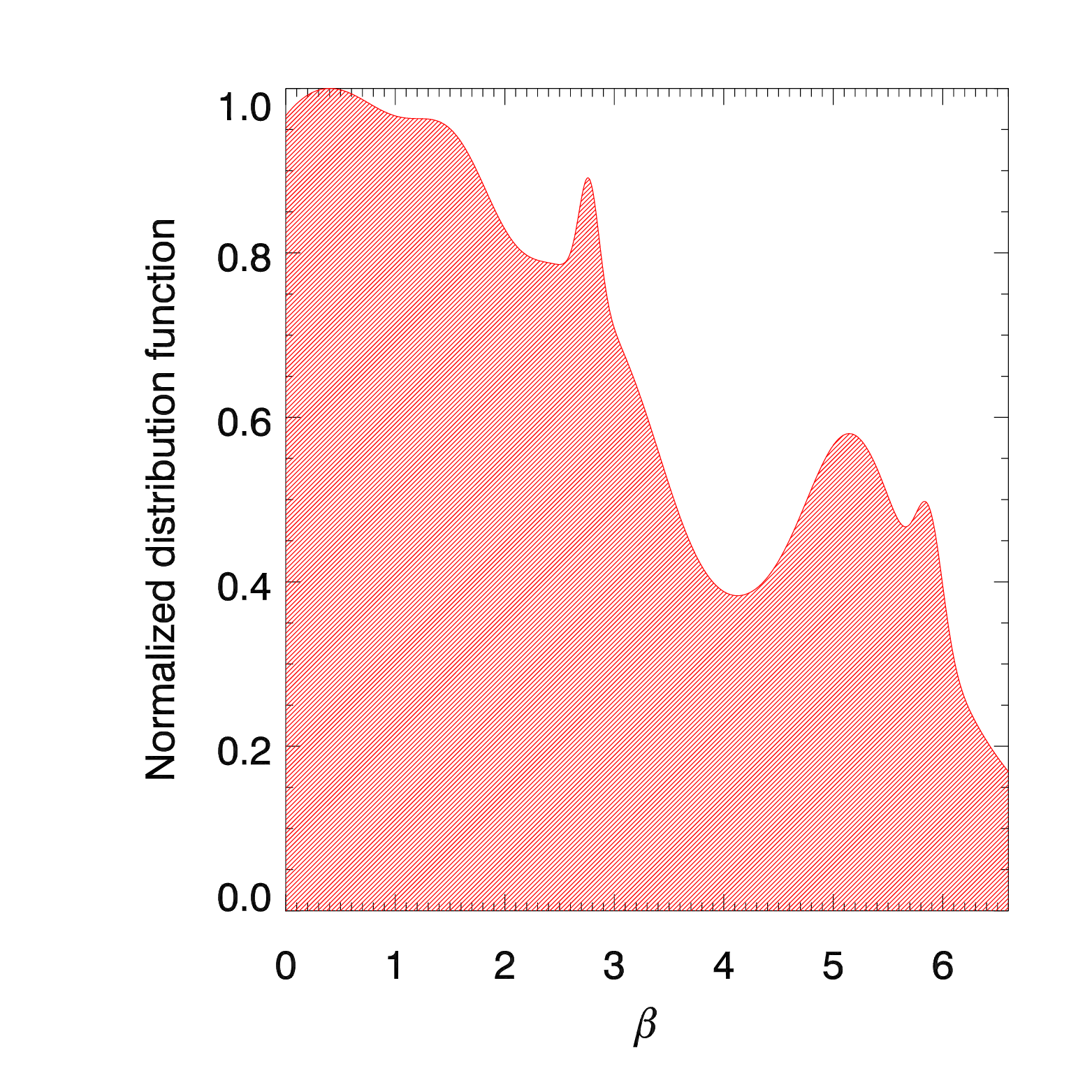}
}
\hbox{
\hspace{-1.0cm}
\includegraphics[width=90mm,height=70mm]{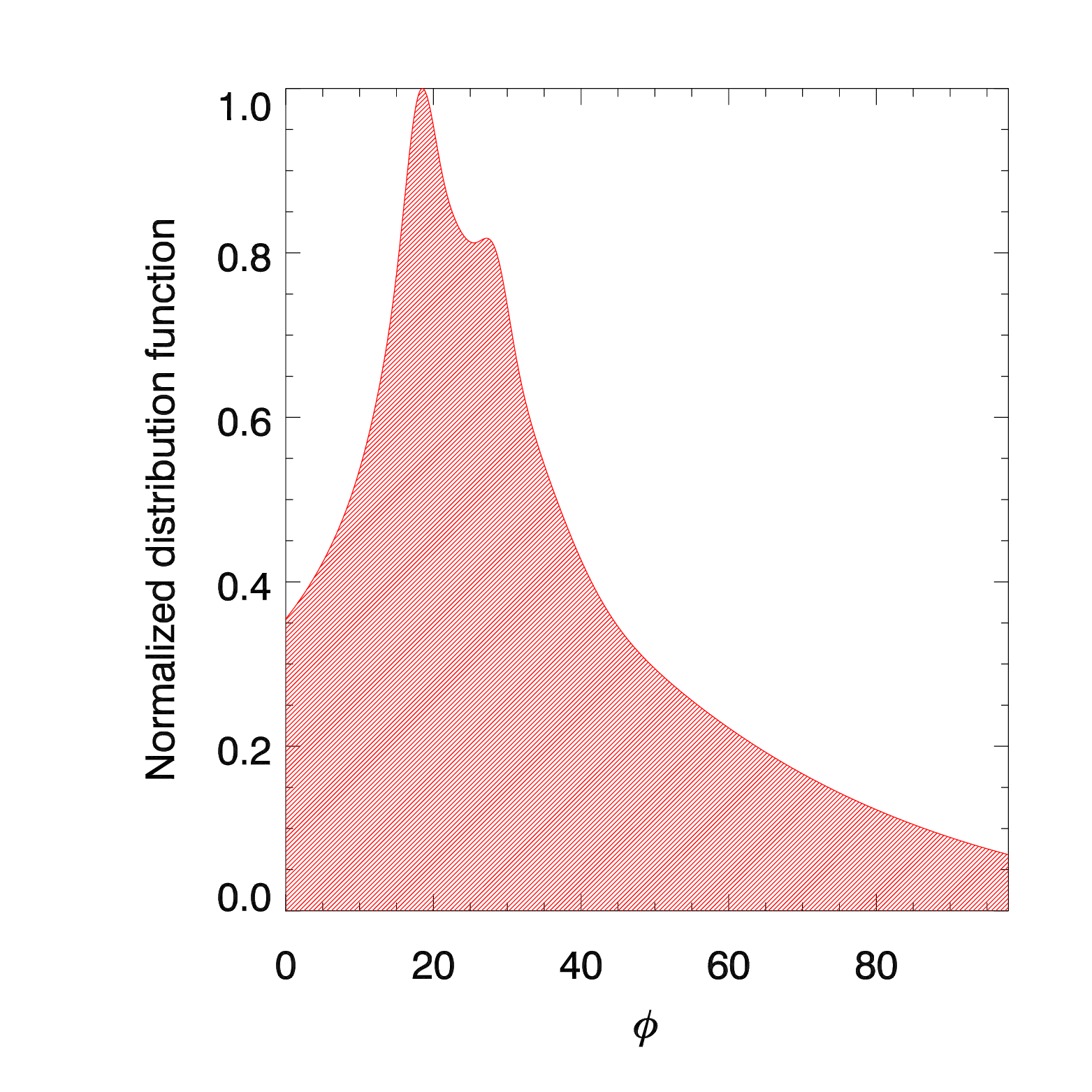}
}
\hbox{
\hspace{-1.0cm}
\includegraphics[width=90mm,height=70mm]{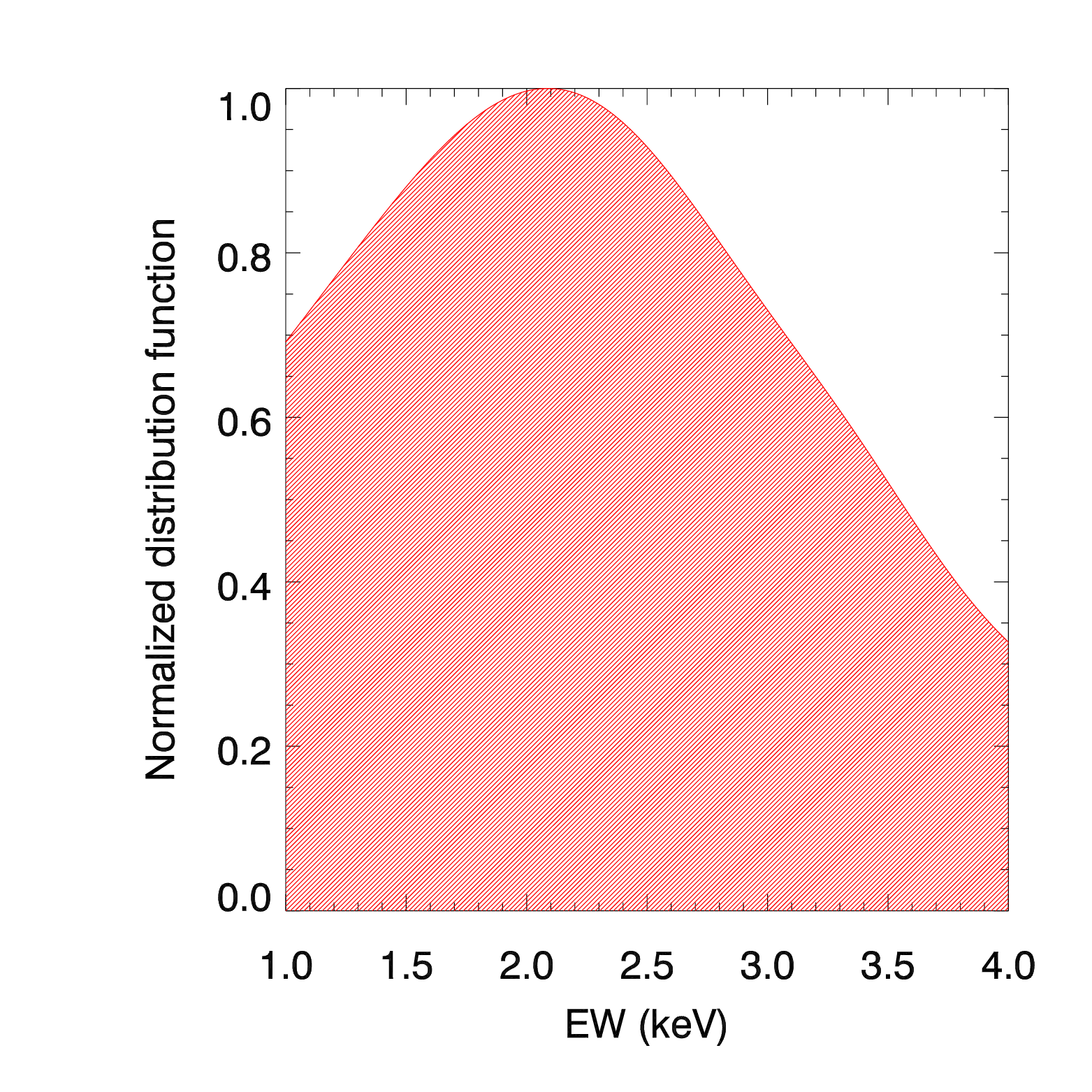}
}
\caption{Normalized distribution functions for $\beta$,
$\phi$ and EW.}
\label{fig4}
\end{figure}

In Fig.~\ref{fig4} normalized distribution functions are shown
for the best-fit
parameters on which the {\tt kyrline} model depends. These distributions
are generated combining one Gaussian distribution for each
individual measurement, whose
mean (standard deviation) is the best-fit
value (statistical error). In Tab.~\ref{tab2} the mean ($\langle \rangle$)
\begin{table}
\begin{center}
\begin{tabular}{lccc} \hline \hline
Parameter & $\langle \rangle$ & $\sigma$ & $N_{obj}$ \\ \hline
$\beta$ & -2.6 & 1.8 & 22 \\
$\phi$ & 33 & 21 & 21 \\
$\log(EW)$ & 2.4 & 1.4 & 20 \\ \hline \hline
\end{tabular}
\end{center}
\caption{Mean and standard deviations on $\beta$, $\phi$,
and $EW$ in our sample. $N_{obj}$ is the number of objects
on which a statistically meaningful constraints on the
parameter can be set, and which have been used to
construct the normalized distributions in Fig.~\ref{fig4}}
\label{tab2}
\end{table}
and standard deviation ($\sigma$)
of the combined distributions are listed, together with
the number of measurements used to construct the
normalized distribution. In 5 objects a
constraints on the black hole spin can be derived
($\langle a \rangle = 0.6$, $\sigma_a = 0.3$).
In two of them (HE1029-1401, $a = 0.85 \pm 0.04$; NGC~3227,
$a < 0.18$) the measurement
is formally inconsistent with a maximally spinning black hole;
in the other three (MCG-6-30-15, NGC~3516, NGC~4395) only a
lower limit on $a$ can be derived.

Several authors have discussed the effect that the spectral curvature
produced by high-density ionized absorbers may have on the detectability
of broad Fe K$_{\alpha}$ profiles (\cite{reeves04,pounds03,turner05}; see as well
Diaz-Trigo's and
Reeves' papers in this volume). If not properly accounted for by
``warm absorbers'' models, this curvature may mimic the red wings
of a relativistic broadened profile in the 3--6~keV energy range.
Approximating an ionized absorber with a cold absorber - as done with
model~(\ref{eqn1}) - might be too crude.

\begin{figure}
\hbox{
\includegraphics[width=75mm,height=70mm]{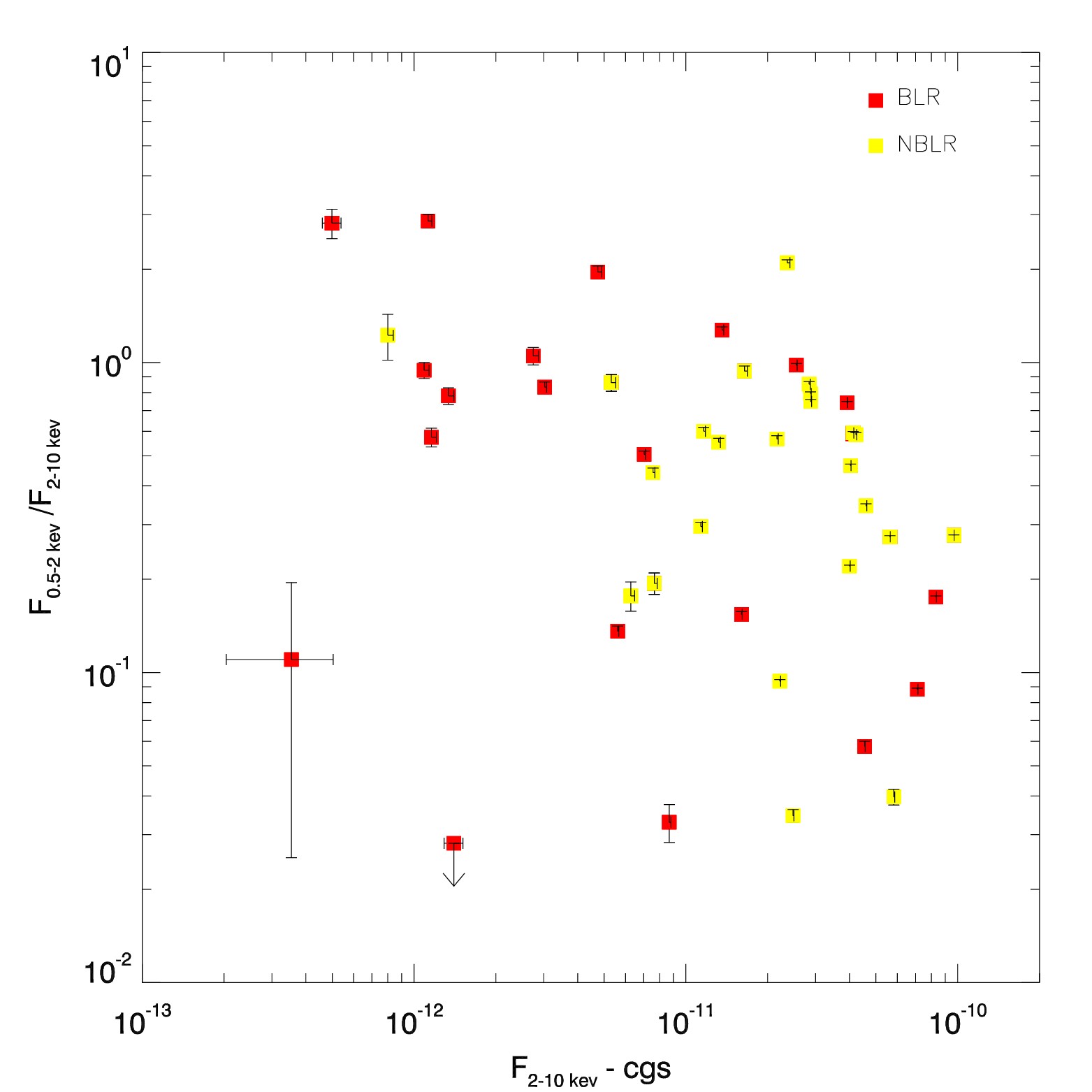}
}
\caption{Softness flux ratio against 2--10~keV flux
for the AGN of our sample, where only an upper limit on the EW
of the iron K$_{\alpha}$ broad component is found ({\it red squares})
and AGN where a broad iron K$_{\alpha}$ line is detected
({\it yellow squares}).}
\label{fig5}
\end{figure}

In Fig.~\ref{fig5} we plot the 0.5--2~keV versus 2--10~keV softness
ratio for the objects of our sample, separating with different colors
objects where only
an upper limit on the broad Fe K$_{\alpha}$ line $\le 100$~eV has been
measured from objects where a broad line has been detected. The softness
ratio should be sensitive to the presence of ``warm absorbers'',
which imprint
strong absorption features in the soft band\footnote{This ratio is actually
mostly sensitive to standard warm absorbers with photoionization parameters
$\xi \sim$10--100 and $\log (N_H) \sim$10$^{21-22}$~cm$^{-2}$
(\cite{blustin05}). However, high-density, high-ionization warm absorber are
often found in conjunction with lower-density, low ionization
counterparts.}.
We find no difference in the softness ratio in the two
classes. We conclude that it
is unlikely that the number of iron K$_{\alpha}$ broadened line
detections is significantly affected by uncertainties in
the underlying warm absorbed continuum determination (although
in individual objects this effect may have an important impact,
as discussed by the above referred authors).

\section{Detection of relativistically broadened profiles in stacked spectra}

Relativistically broadened iron K$_{\alpha}$ profiles have been detected
in about 25\% of the sources of our sample. In at least 50\% of the
sample, the upper limits on the EW of a
relativistically broadened profile are
inconclusive as to establishing whether such a feature is present
(cf. Fig.~\ref{fig2} and \ref{fig3}).

\begin{figure*}
\hbox{
\includegraphics[width=75mm,height=70mm]{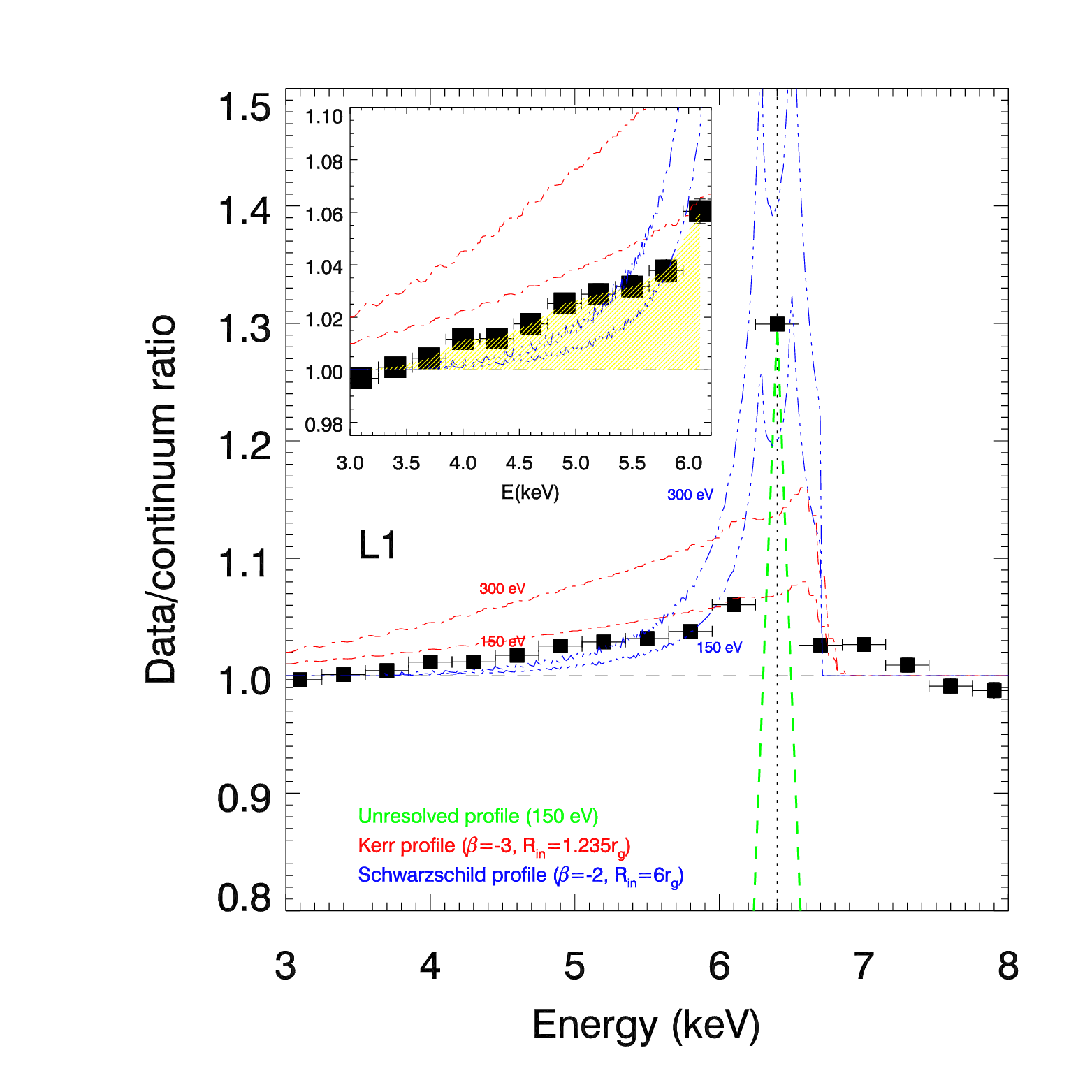}
\includegraphics[width=75mm,height=70mm]{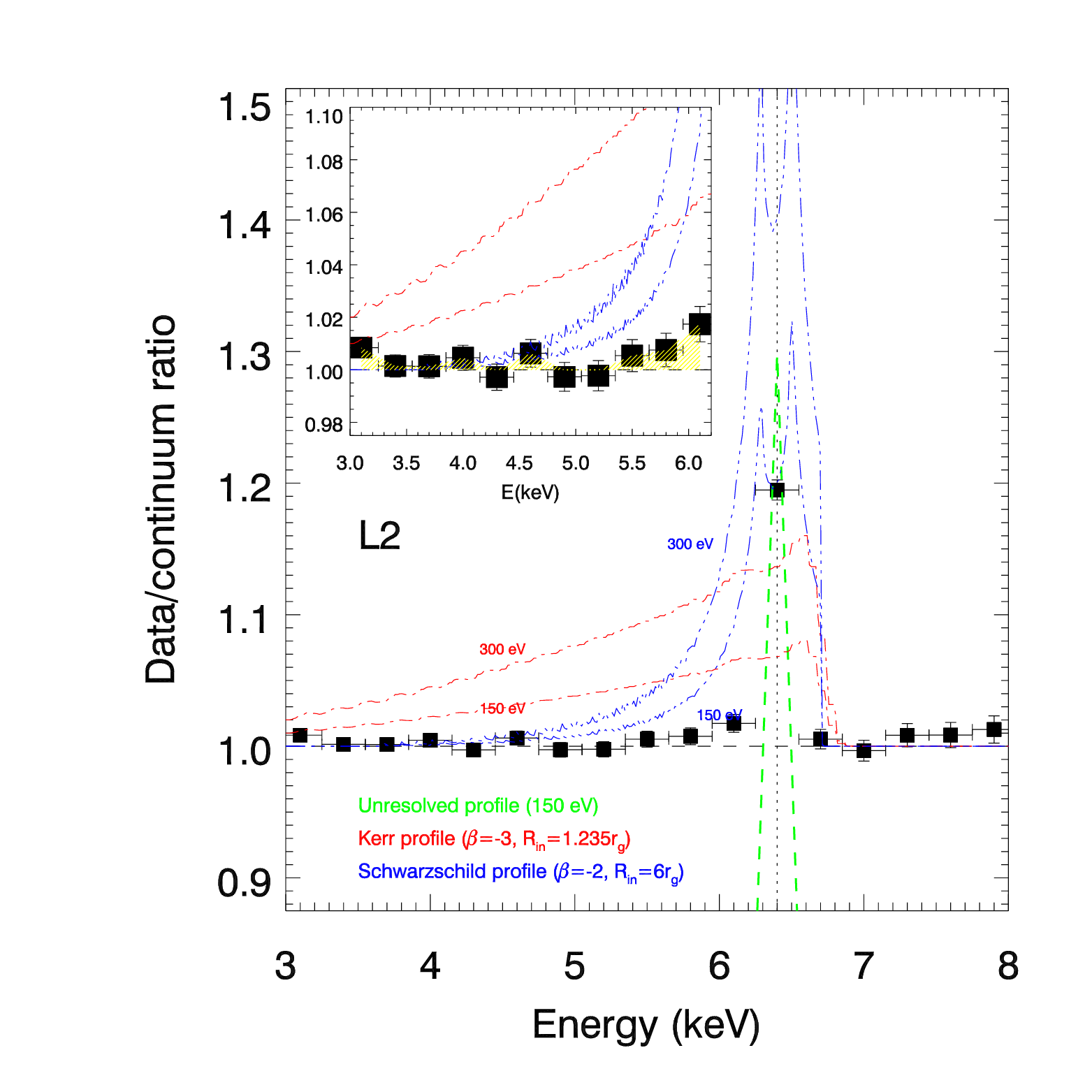}
}
\hbox{
\includegraphics[width=75mm,height=70mm]{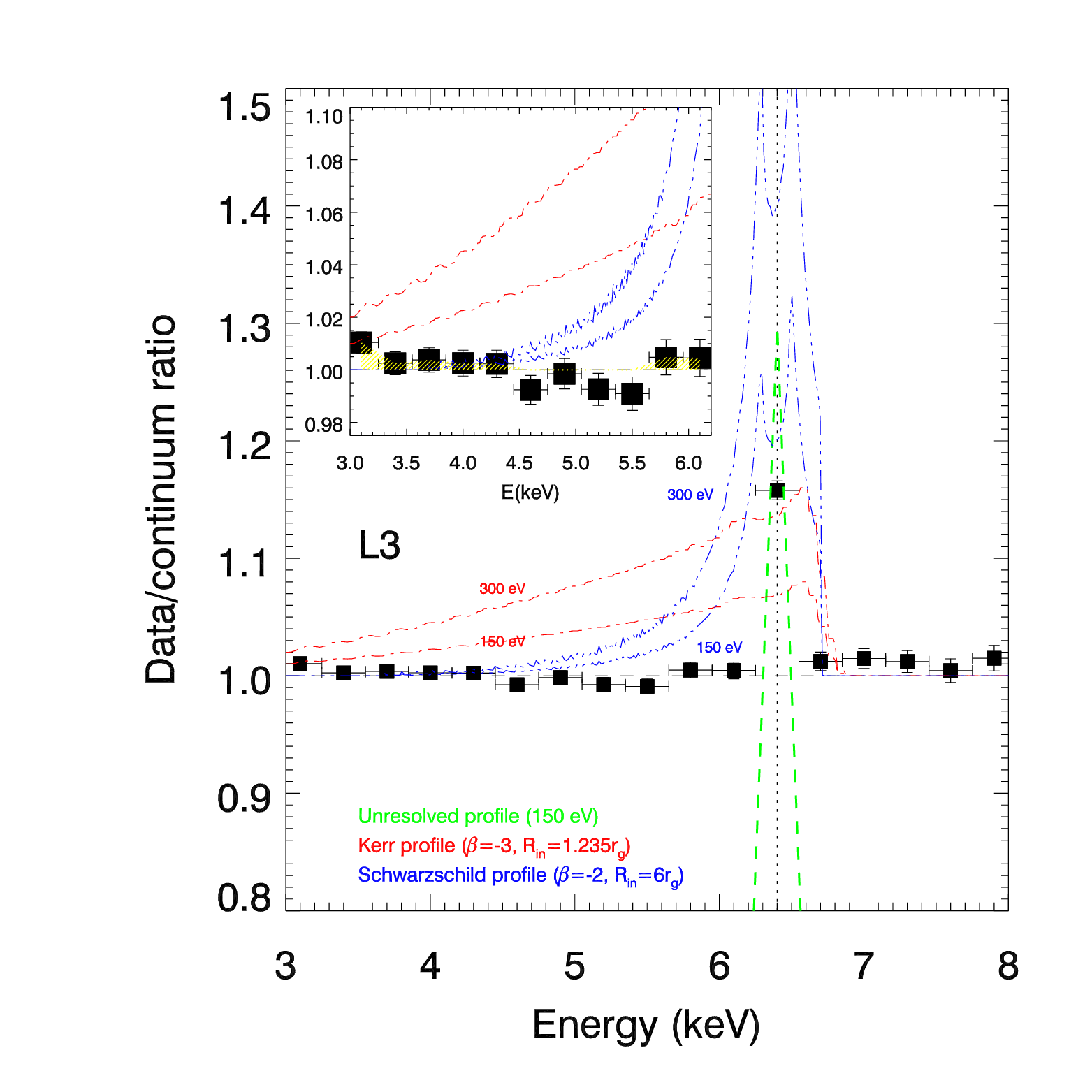}
\includegraphics[width=75mm,height=70mm]{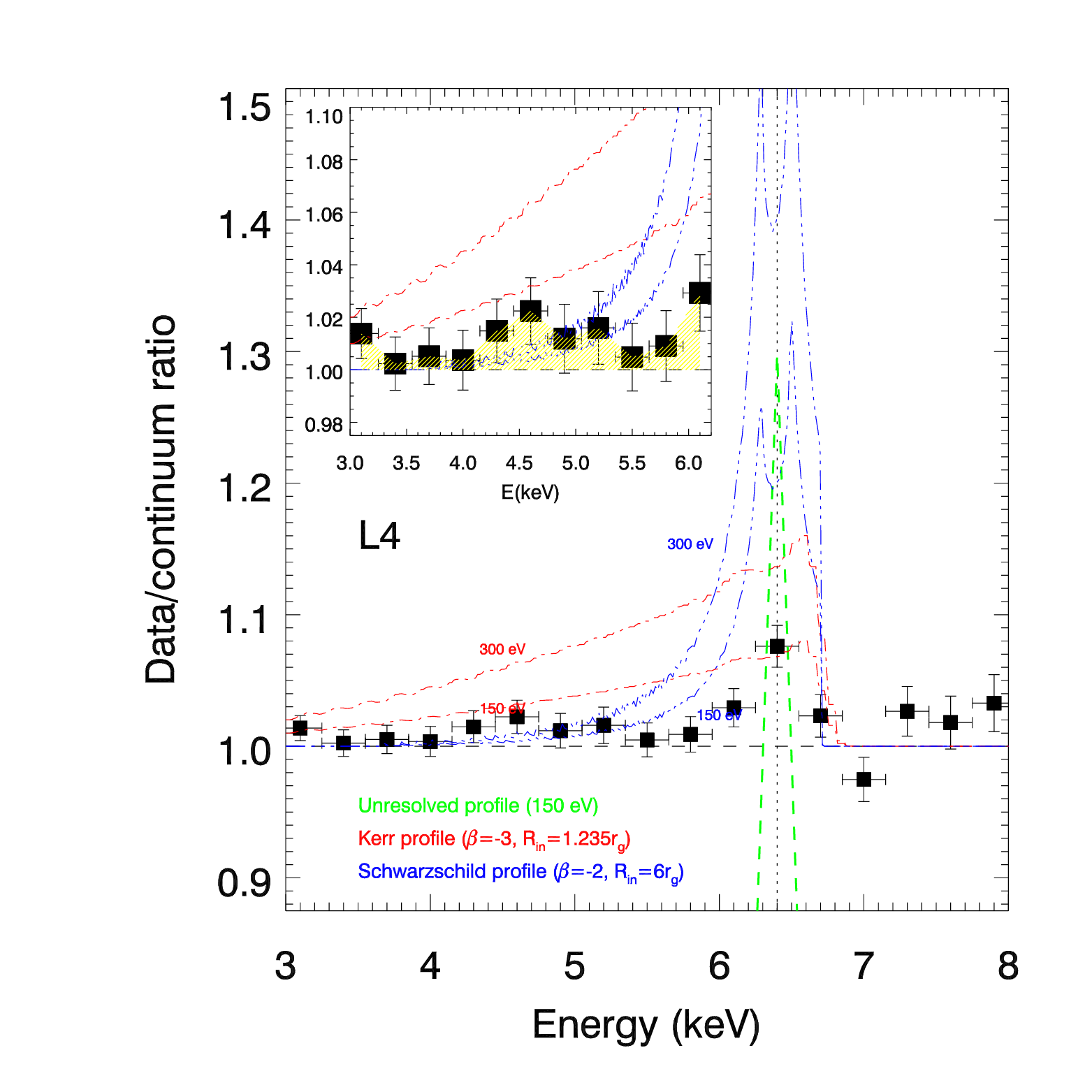}
}
\caption{3--8~keV (3--6~keV in the {\it insets})
stacked residuals for the AGN of our sample, divided
in four equally-populated 2--10~keV intrinsic
luminosity bins. The curves represent the: a) expected
profile of an unresolved Gaussian profile, reflecting
the intrinsic pn camera energy resolution ({\it
dashed line}); b) two maximally rotating
Kerr profiles corresponding to EW=150~eV and
300~eV ({\it dot-dashed profiles}); c) two Schwarzschild
profiles ({\it long dashed-dot line}) corresponding to the
same values of EW. The luminosity bins
are defined as follows: L1, $L_X < 10^{43}$~erg~s$^{-1}$;
L2, $10^{43} \le L_X < 5 \times 10^{43}$~erg~s$^{-1}$;
L3, $5 \times 10^{43} \le L_X < 1.5 \times 10^{44}$~erg~s$^{-1}$;
L4, $L_X \ge 1.5 \times 10^{44}$~erg~s$^{-1}$}
\label{fig6}
\end{figure*}

\begin{figure*}
\hbox{
\includegraphics[width=75mm,height=70mm]{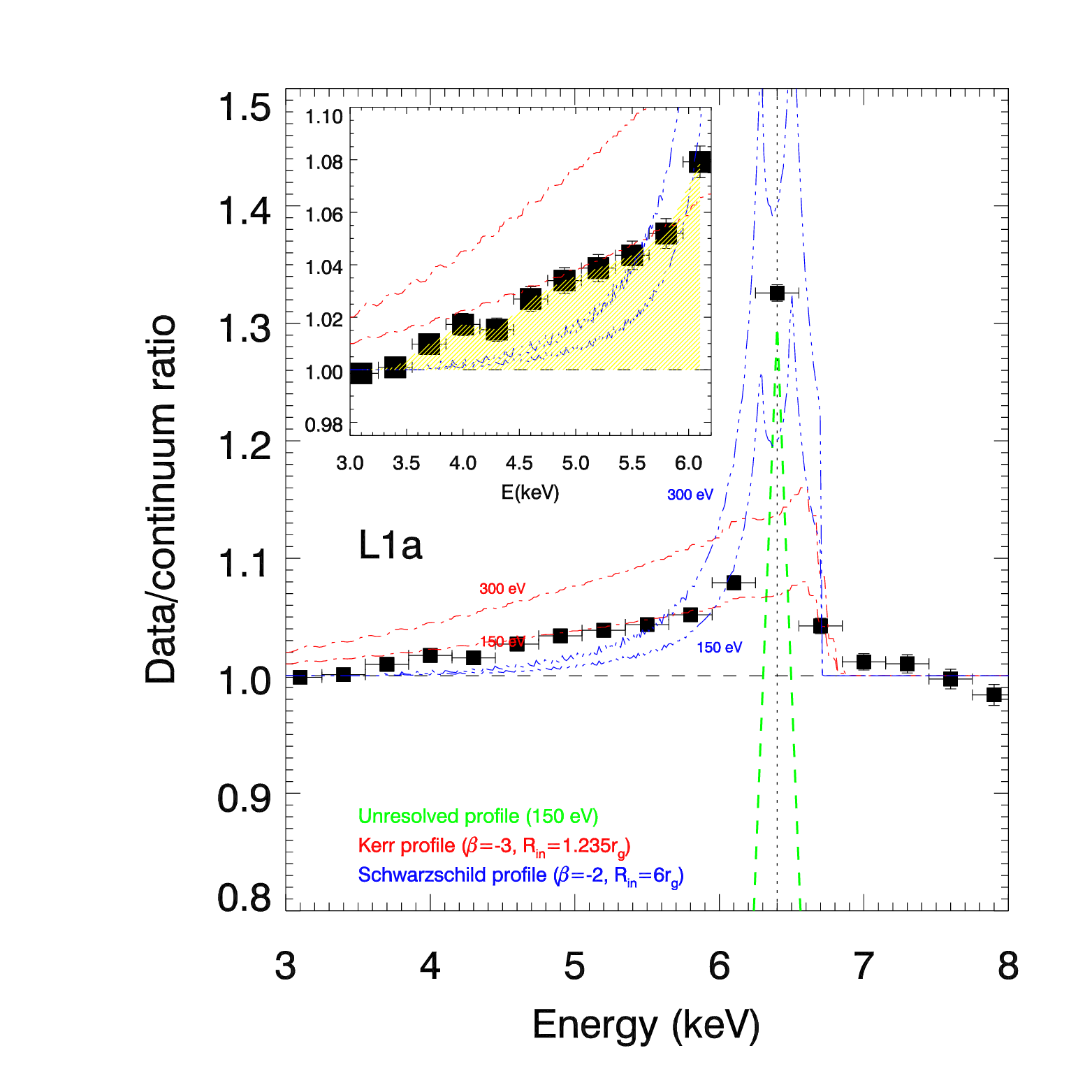}
\includegraphics[width=75mm,height=70mm]{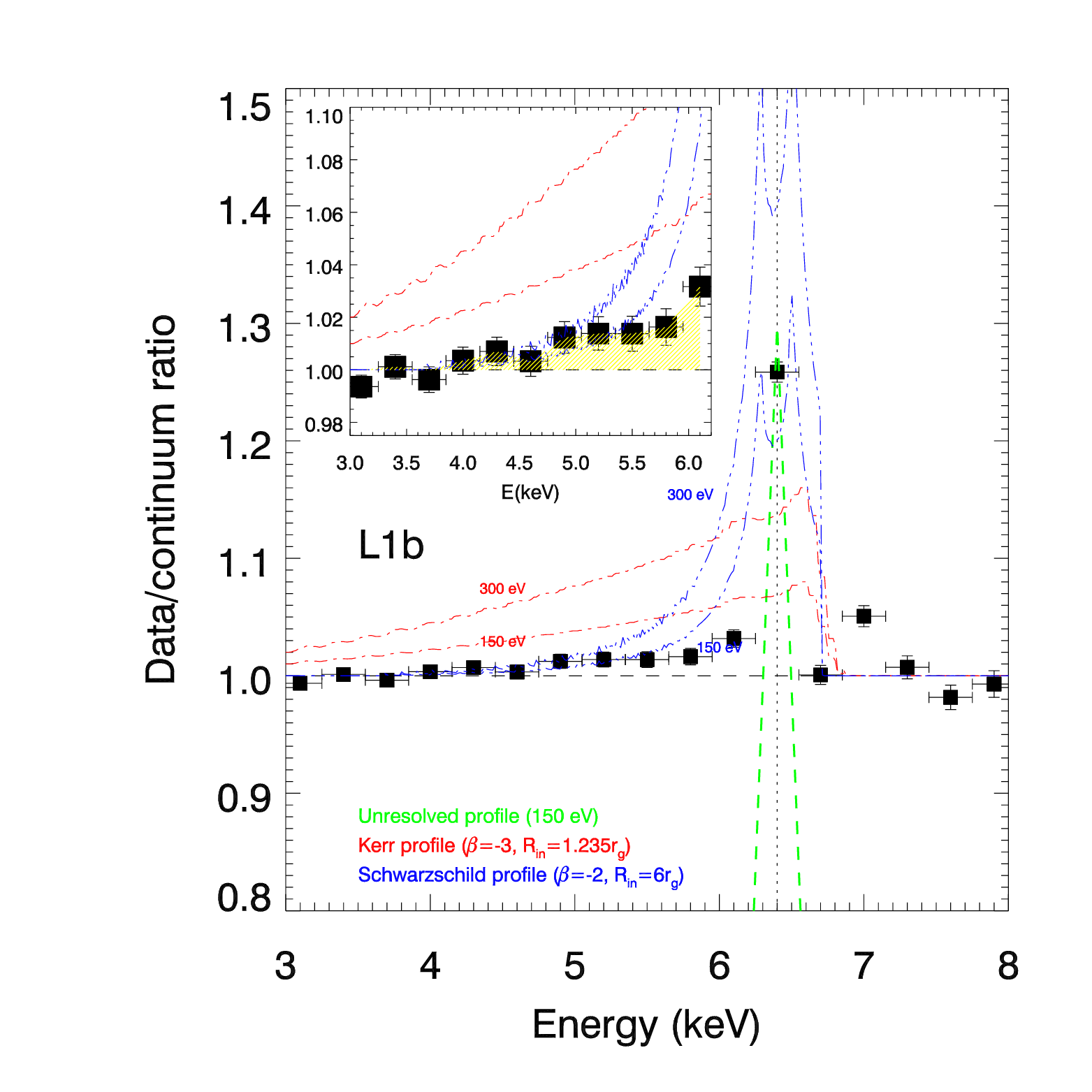}
}
\caption{The same as Fig.~\ref{fig6}, when the lowest-luminosity
bin is split in two equally-populate (in source numbers) bins.
The {\it left panel} corresponds to objects with
$L_X \le 10^{42}$~erg~s$^{-1}$, the {\it right panel} to
$10^{42} < L_X \le 10^{43}$~erg~s$^{-1}$.}
\label{fig7}
\end{figure*}

This statistical limitation, however, can be overcome, by stacking data together.
In Fig.~\ref{fig6} we show stacked residuals for the objects of our sample,
once they have been divided in four, equally populated (as of number of
sources) 2--10~keV
absorption-corrected luminosity classes. The stacked residuals were generated
by summing together the residuals of a fit with model~(\ref{eqn1}), after the
$\Sigma_i G_i$ and $B(\beta, \phi, a)$ terms had been removed.
A clear
broadened relativistic profile is present in the lowest luminosity
sub-sample ($L_X < 10^{43}$~erg~s$^{-1}$),
and possibly in the highest luminosity sub-sample.
If the lowest luminosity bin
is split into two equally-populated bins,
the broad line keeps being stronger in the
lower-luminosity bin (Fig.~\ref{fig7}).
It seems to exist a sort of ``Baldwin effect'' on
the relativistically broadened
component of the iron K$_{\alpha}$, similar to what already
claimed on the narrow component of the same
line (\cite{iwasawa93,page03}; see as well
Jim\'enez-Bail\'on et al. 2005 for a different
interpretation of the same data).

\begin{figure*}
\hbox{
\includegraphics[width=75mm,height=70mm]{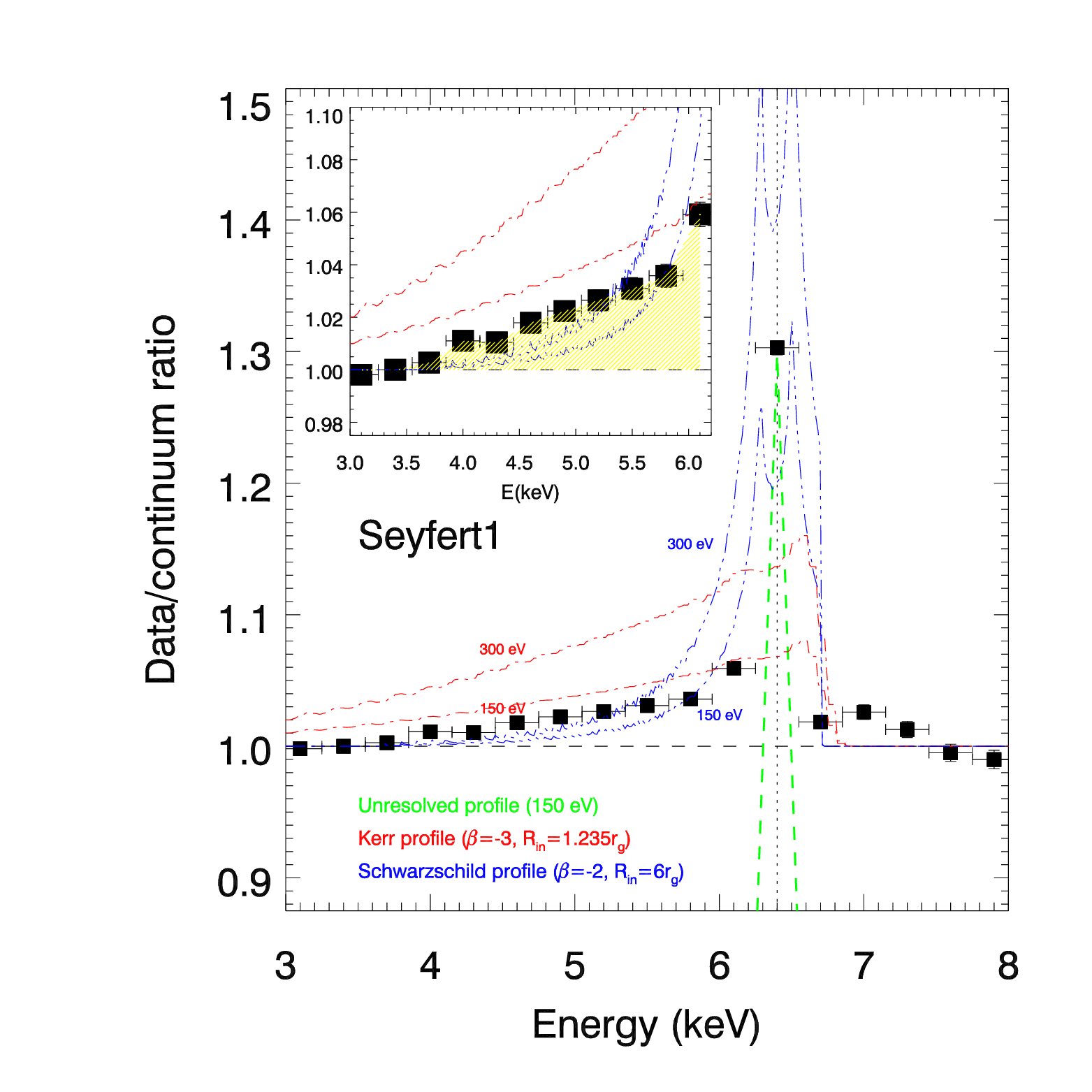}
\includegraphics[width=75mm,height=70mm]{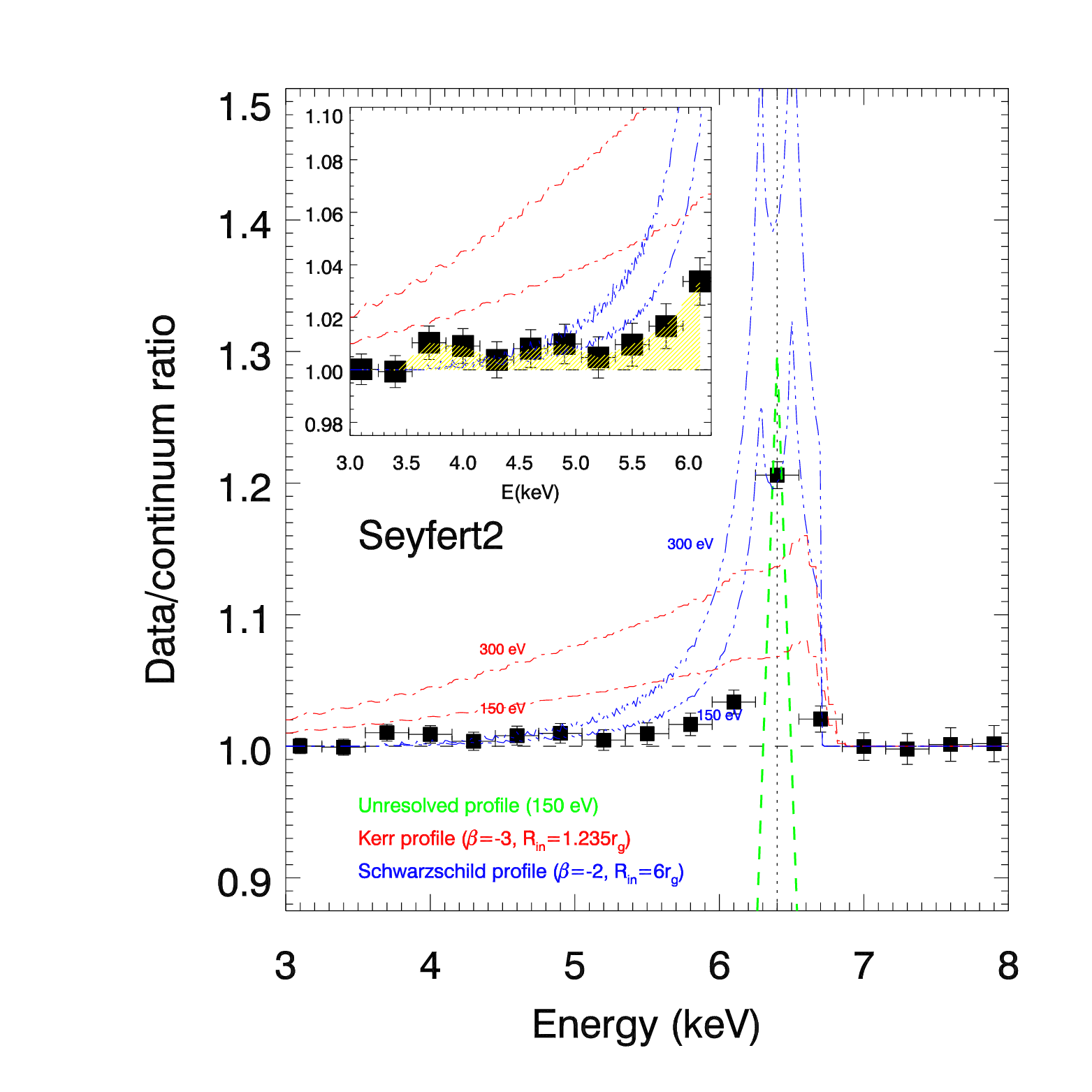}
}
\caption{The same as Fig.~\ref{fig6}, for luminosity-matched
sub-samples of Seyfert~1s ({\it left panel}) and
Seyfert~2s ({\it right panel}).}
\label{fig8}
\end{figure*}

Broadened profiles with comparable profiles
are detected in the stacked residuals
of luminosity-matched samples of Seyfert~1s and Seyfert~2s 
(see Fig.~\ref{fig8}),
although in the latter the profile is weaker.
The integrated EW of the
stacked profiles is $\approxlt$150~eV for any
luminosity class (compare the
observed profiles with the profiles
of an extreme Kerr line in Fig.~\ref{fig6} to
Fig.~\ref{fig8}), therefore significantly
lower than measured by Streblyanska et al. (2005) on the Lockman
Hole AGN.

\section{Can we achieve the full coverage of a complete sample?}

The paradigm to explain the origin of the energy output in
Active Galactic Nuclei (AGN) requires radiatively efficient
accretion onto super-massive black holes (Lynden-Bell 1969).
In this scenario, the
accretion flow reaches the innermost regions close to the black
hole, possibly the innermost circular stable orbit if it occurs
in a standard geometrically thin and optically thick disk.
An inescapable prediction of this scenario is that relativistic
broadening and skewing of emission lines should be common in AGN.
The measure of the ``true''
fraction of objects where relativistic effects
are detected would be crucial to establish if
and how the standard AGN paradigm needs to be modified.
Current AGN samples in X-rays are inevitably biased towards bright
AGN, and furthermore towards bright Seyferts where the detection
of a relativistically broadened iron K$_{\alpha}$ profile was
expected on the basis of observations prior to the {\it Chandra}
and XMM-Newton launches.

The AGN sample on which the study presented in this paper has been conducted
is neither complete nor unbiased in any statistical sense. It is merely the
collection of all the observations available in the XMM-Newton science
archive at a given moment in time.
This study needs to be extended (or restricted) to a
statistically complete flux-limited sample before any
inference can be derived on the true fraction of AGN where relativistic
effects are important, as well as
on the properties of the innermost regions of the accretion flow around
super-massive black holes (or, hopefully, on the black hole spin
itself). The double bias outlined above needs to be removed.

How can one build a complete sample, where the spectrum of each
member of the sample has enough statistics for
meaningful constraints on the properties of relativistically broadened iron
K$_{\alpha}$ lines to be derived?
Let's consider again the EW versus net counts correlation in Fig.~\ref{fig9}.
For counts $\approxgt 10^5$ the correlation flattens. This represents
the minimum number of net counts necessary to detect a line as strong as the
average of the EW distribution. Moreover, a number of net
counts $\approxgt 2 \times 10^5$ ensures that a complete separation
between measurements and upper limits is achieved. This condition is crucial to
assess the significance of any non detections.
Hence, we conclude that in order to derive the true fraction of relativistic
lines in a sample of AGN, each of the members of the
sample has to be observed long enough to
accumulate at least 200~000 net counts in the hard X-ray band.

If we take the Piccinotti sample (\cite{piccinotti82}) as example of a
flux-limited sample in the hard X-ray band,
only 6 out of its 27 type~1 objects have
been optimally exposed with XMM-Newton.
The detection fraction of relativistically
broadened iron K$_{\alpha}$ lines in this sub-sample is $50 \pm 32 \%$.
We estimate that a full coverage of a hard X-rays
flux limited sample
such as the Piccinotti or the RXTE All Sky Survey
(\cite{revnivtsev04}) would require $\sim$1~Ms
at a flux limit $\simeq$$2 \times 10^{-11}$~erg~cm$^{-2}$~s$^{-1}$. This
is large but not unreasonable time allocation for an XMM-Newton
Large Program.

\section{Conclusions}

The main conclusions of this study can be summarized as follows:

\begin{itemize}

\item relativistically broadened Fe K$_{\alpha}$ lines are present in
$\simeq$50\% of well exposed XMM-Newton spectra

\item the detection rate and average properties
are not significantly different between type~1 and type~2
objects, although the analysis of stacked spectral residuals
on luminosity well-matched samples of Seyfert~1s and
Seyfert~2 suggests a possible weaker line in the latter

\item the relativistically broadened Fe K$_{\alpha}$ lines
detected in our study
cover a large range of EW. However, our
stacked spectral residuals analysis
finds typically EW$\approxlt$150~eV, far from the $\simeq$500~eV EWs claimed
by Streblyanska et al. (2005)
for the Lockman Hole AGN

\item relativistically broadened Fe K$_{\alpha}$ lines
are significantly more common in low luminosity
($L_X \le 10^{43}$~erg~s$^{-1}$) AGN

\item the average
disk inclination is $\langle \phi \rangle = 34^{\circ}$;
the average accretion flow power-law radial dependent index is $\langle \beta \rangle = -2.7$

\item in 4 out of 6 objects where the measurement
of the dimensionless black hole spin yields meaningful
constraints, the Schwarzschild solution can be ruled out

\end{itemize}

The conclusions of this study are still based on an incomplete and biased
sample. However, a relatively modest investment of XMM-Newton time could
allow them to be put on a firm statistical ground. {\bf At
least 200~000 hard X-ray counts are necessary to ensure an unambiguous
detection}, as well as {\bf a clear discrimination between detections and upper
limits.}
We estimate that around 1~Ms
would be enough to optimally expose a hard X-ray flux limit
sample at a flux limit $\simeq$$2 \times 10^{-11}$~erg~cm$^{-2}$~s$^{-1}$.

\end{document}